\documentclass[11pt,a4paper,reprint,notitlepage,superscriptaddress,prd]{revtex4-1}

\usepackage{amsmath,amssymb,amsfonts}

\usepackage{physics}

\usepackage[dvipsnames]{xcolor}
\usepackage{graphicx}

\usepackage[pdftex]{hyperref}
\hypersetup{
    colorlinks=true,   
    linkcolor=Blue, 
    citecolor=purple, 
    filecolor=magenta
}

\usepackage[capitalise]{cleveref}

\renewcommand{\t}[1]{\mathrm{{#1}}}

\begin{document}

\title{Unruh Effect of Detectors with Quantized Center-of-Mass}

\author{Vivishek Sudhir}
\thanks{vivishek@mit.edu}
\affiliation{Department of Mechanical Engineering, Massachusetts Institute of Technology, Cambridge, MA 02139, USA}
\affiliation{LIGO Laboratory, Massachusetts Institute of Technology, Cambridge, MA 02139, USA}

\author{Nadine Stritzelberger}
\affiliation{Department of Applied Mathematics, University of Waterloo, Waterloo, ON N2L 3G1, Canada}
\affiliation{Institute for Quantum Computing, University of Waterloo, Waterloo, ON N2L 3G1, Canada}
\affiliation{Perimeter Institute for Theoretical Physics,
Waterloo, ON N2L 2Y5, Canada}

\author{Achim Kempf}
\affiliation{Department of Applied Mathematics, University of Waterloo, Waterloo, ON N2L 3G1, Canada}
\affiliation{Institute for Quantum Computing, University of Waterloo, Waterloo, ON N2L 3G1, Canada}
\affiliation{Perimeter Institute for Theoretical Physics,
Waterloo, ON N2L 2Y5, Canada}
\affiliation{Department of Physics, University of Waterloo, Waterloo, ON N2L 3G1, Canada}

\begin{abstract}
The Unruh effect is the prediction that particle detectors accelerated through the vacuum get excited 
by the apparent presence of radiation quanta --- a fundamental quantum phenomenon 
in the presence of acceleration. 
Prior treatments of the Unruh effect, that presume a classically prescribed trajectory, do not account for the 
quantum dynamics of the detector's center-of-mass.
Here, we study more realistic detectors whose center of mass is a quantized degree of freedom being accelerated by 
an external classical field. We investigate the detector's recoil due to the emission of Unruh quanta. 
Vice versa, we also study the recoil's impact on the emission of Unruh quanta and the excitation of the detector. 
We find that the recoil due to the emission of Unruh quanta may be a relevant experimental signature of the Unruh effect.

\end{abstract}
\maketitle

\section{Introduction}

An idealized particle detector --- a two-level system --- accelerated through the vacuum of a quantum field can become 
excited \cite{Unruh_UDW_detector, DeWitt_UDW_detector, Gibbons_Hawking, birrell_davies,CrispMats08}. In the case of uniform acceleration, the
state of the detector assumes a thermal form with an apparent temperature proportional to the acceleration. 
This is the Unruh effect, and the particle detector is eponymously referred to as the Unruh-DeWitt 
(UDW) detector \cite{Unruh_UDW_detector,DeWitt_UDW_detector}.
In the usual treatment, the energy and momentum required to excite the detector, along with the subsequent emission of a photon, come from an
unspecified external agent which 
enforces a prescribed classical uniformly accelerated trajectory for the detector. 
These idealized treatments of the Unruh effect neither describe the accelerating agent, nor the recoil of the detector
due to photon emission.
Both are important, however, 
to the question of the viability of an experimental observation of the Unruh effect. 
For example, real experiments cannot sustain uniform acceleration indefinitely, and also detector recoil makes it unrealistic to consider uniform acceleration. 
In fact, the recoil could itself be an avenue towards the detection of the Unruh effect. Importantly, however, in this case
the sensitivities required to resolve the recoil are likely to be comparable to the quantum fluctuations of the
detector's center-of-mass motion. This means that the recoil needs to be calculated within a full quantum mechanical treatment of the detector's center-of-mass degree of freedom.
From this perspective, the couple of works studying detector recoil from the Unruh effect (e.g., \cite{Paren95,Casadio_Venturi_1,Reznik_1998}) remain incomplete.

Here, we dynamically account for the acceleration of the detector, and self-consistently treat the
recoil of the detector center-of-mass.
To this end, the detector's center of mass degrees of freedom are treated quantum mechanically
\cite{Stritzelberger_Kempf},
and its acceleration is described by  coupling the detector to an external classical accelerating field. 
Within this framework, we study the vacuum excitation process for the internal and center of mass degrees of freedom
of the detector --- a process we term the \emph{massive Unruh effect}, in reference to the detector mass being assumed finite. By letting the detector mass go to infinity, the behavior of classical detector trajectories can be recovered. We also compute the recoil pattern and its relationship to the pattern of the emission of Unruh quanta. Both patterns assume characteristic non-isotropic forms. 
The peak emission probability is proportional to the acceleration, which allows us to informally 
talk of a ``temperature'' for the massive Unruh effect, in analogy with the conventional Unruh effect.

\section{Unruh effect with infinite-mass detector}
\label{Sec:Unruh_effect_UDW}

We briefly recall the Unruh effect for a detector accelerated on a prescribed trajectory through a scalar quantum field. The detector-field system is described by the free Hamiltonian
\begin{equation}
	\hat{H}_0 =
     \Omega \ket{e}\bra{e} 
     + \int d^3 k \, ck \, \hat{a}_{\mathbf{k}}^{\dagger} \hat{a}_{\mathbf{k}}^{\vphantom \dagger},
\end{equation}
where $\Omega$ denotes the energy gap of the detector's ground ($\ket{g}$) and excited ($\ket{e}$) states, and 
$\hat{a}_{\mathbf{k}}^{\dagger}$ ($\hat{a}_{\mathbf{k}}^{\vphantom \dagger}$) is the creation (annihilation) operator 
of the scalar quantum field mode of momentum $\mathbf{k}$.
Following Unruh and DeWitt \cite{Unruh_UDW_detector,DeWitt_UDW_detector}, and in analogy with realistic models of 
light-matter interaction in electrodynamics \cite{Cohen}, 
we assume that the detector's internal state couples to the field through the interaction Hamiltonian,
\begin{equation}
	\hat{H}_\t{int} = q\, \hat{\mu}(t) \otimes \hat{\phi}(\mathbf{x}(t)).
\end{equation}
Here, $q$ is the coupling strength,
$\hat{\mu}=\ket{e}\bra{g}+\ket{g}\bra{e}$ is the detector's ``monopole'' moment, and $\hat{\phi}(\mathbf{x}(\tau))$ is
the scalar field operator along the detector's trajectory.  
We limit ourselves to the regime of non-relativistic detector velocity, which allows us to identify the detector's proper 
time $\tau$ with the coordinate time $t$.

The crux of the Unruh effect is that the detector can be excited by accelerating it through the quantum vacuum of the 
scalar field. 
Given the structure of the interaction, the detector can be excited only if the quantum field is simultaneously excited to (at least)
the single particle quantum state, $\hat{a}^{\dagger}_{\mathbf{k}} \ket{0}$.
That is, contrary to resonance effects such as absorption, the Unruh effect is the result of counter-rotating-wave 
terms in $\hat{H}_\t{int}$ \cite{Scully18}.
To elucidate this, we compute the probability that the initial state,
$\ket{\psi_\t{i}}=\ket{g}\otimes \ket{0}$ -- the joint ground state of the system -- transitions to the final state,
$\ket{\psi_\t{f}}=\ket{e}\otimes \hat{a}^{\dagger}_{\mathbf{k}} \ket{0}$, where both the detector and the field are excited.
Working in the interaction picture defined by the free Hamiltonian, in which,
\begin{align*}
	\hat{\mu}(t) &= e^{i \Omega t}\ket{e}\bra{g} + \t{h.c.} \\
	\hat{\phi}(\mathbf{x},t) &= \int \frac{d^3 k}{(2\pi)^{3/2}} \sqrt{\frac{ c^2}{2k}} 
		\left[ e^{-ick t + i\mathbf{k}\cdot\mathbf{x} }\hat{a}_{\mathbf{k}}+ \t{h.c.} \right],
\end{align*}
the probability amplitude (to first order in perturbation theory) for the afore-mentioned excitation process is,
\begin{equation*}
    \mathcal{A}_\t{U}(\mathbf{k}) = \bra{\psi_\t{f}}\int_{-\infty}^{\infty} dt \,\hat{H}_\t{int}(t) \ket{\psi_\t{i}}
    = \frac{qc}{2\pi \sqrt{4\pi k}} \,\mathcal{I},
\end{equation*}
where, 
\begin{equation}\label{eq:J}
    \mathcal{I} = \int_{-\infty}^{\infty} dt\, 
    e^{it(ck+\Omega) - i \mathbf{k}\cdot \mathbf{x}(t)}.
\end{equation}
We now define the probabilities for: 
the detector to get excited and a field quantum of momentum $\mathbf{k}$ to be emitted,
\begin{equation*}
	P_\t{U}(\mathbf{k}) = \abs{\mathcal{A}_\t{U}(\mathbf{k})}^2;
\end{equation*}
and the excitation of the detector, irrespective of the momentum of the emitted photon,
\begin{equation*}
	P_\t{U} = \int d^3 k \, \abs{ \mathcal{A}_\t{U}(\mathbf{k}) }^2.
\end{equation*}

When the detector is in inertial motion, $\mathbf{x}(t)= \mathbf{x}_0+\mathbf{v}t$,
the excitation amplitude is zero:
\begin{align*}
	\mathcal{A}_\t{U} &= 
		\frac{qc}{2\pi \sqrt{4\pi k}} \,
    	\int_{-\infty}^{\infty} dt\, 
    	e^{it(ck+\Omega-\mathbf{k}\cdot\mathbf{v}) - i \mathbf{k}\cdot \mathbf{x}_0} \\
    &= \frac{qc}{\sqrt{4\pi k}} \,e^{- i \mathbf{k}\cdot \mathbf{x}_0} \delta(\Omega+ck-\mathbf{k}\cdot\mathbf{v}) \\
    &= 0,
\end{align*}
essentially because $\Omega +ck -\mathbf{k}\cdot \mathbf{v} \neq 0$, owing to the fact the energy gap is positive
($\Omega > 0$), the photon momentum is positive ($k>0$), and the detector speed is less than that of 
light ($\abs{\mathbf{v}}< c$).
That is, in inertial motion through the vacuum, the detector does not get excited.

In contrast, consider the detector in non-inertial motion with a uniform acceleration $a$ for time $T$ along
the $z-$direction, i.e.,
\begin{equation*}
    \mathbf{a}(t) = a\, \Theta(t)\Theta(T-t)\, \mathbf{e}_z,
\end{equation*}
where $\Theta$ is the Heaviside step function.
Assuming that the detector's initial position coincides with the origin of the coordinate system, its 
trajectory is,
\begin{equation}\label{eq:accelerated_prescribed_trajectory}
	\mathbf{x}(t) = \left[ \frac{a t^2}{2} \Theta(t) \, \Theta(T-t)
		+ \frac{aT}{2} \left(2t - T \right) \Theta(t-T)
	 \right] \mathbf{e}_z.
\end{equation}
In the following, we restrict the time duration $T$ so that the velocity developed in that time with
an acceleration $a$ is well within the non-relativistic regime;
in particular, we will always take, $|\mathbf{v}(T)| = a T \lesssim 0.01 c$.
Within this non-relativistic regime, and
for the spatial trajectory in \cref{eq:accelerated_prescribed_trajectory}, 
we obtain for the time integral given in \cref{eq:J} [here we define, $\omega = \Omega + ck$,
$\omega' = \omega - a k_z T$, and, $k_z$ the $z-$component of the momentum of the emitted Unruh 
photon, $\mathbf{k}=(k_x,k_y,k_z) = (k\sin(\theta)\cos(\phi) , k\sin(\theta)\sin(\phi) , k\cos(\theta))$],
\begin{equation*}\label{eq:I_Unruh}
\begin{split}
	\mathcal{I} &= 
    	\frac{1}{i \omega} + \pi \delta(\omega)
    		+ e^{iT \left(\omega' + a k_z T/2 \right)} \left( \frac{i}{\omega'} 
    		+ \pi \delta(\omega')\right) \\
    	&\quad + \sqrt{\frac{\pi e^{\frac{i\omega^2}{a k_z}}}{2 i a k_z}} 
    		\left[ \t{Erf}\left( \frac{i \omega}{\sqrt{2i a k_z}} \right)
    			-\text{Erf}\left( \frac{i \omega'}{\sqrt{2i a k_z}} \right) \right]
\end{split}
\end{equation*}
Since both the energy gap of the detector and the absolute value of the momentum of the emitted photon are strictly 
positive ($\Omega>0$ and $k>0$), the delta distribution $\delta(\omega)$ can be omitted. 
Similarly, since the detector's velocity is strictly smaller than the speed of light, we find $\omega' >0$, and so 
$\delta(\omega')$ can also be omitted. Thus,
\begin{equation}\label{eq:I_Unruh_1}
\begin{split}
	\mathcal{I} = \frac{1}{i \omega} &- \frac{e^{iT (\omega' + ak_z T/2)}}{i\omega'} \\
		&+ \sqrt{\frac{\pi e^{i \omega^2/a k_z}}{2 i a k_z}} 
			\text{Erf}\left( \frac{i \omega}{\sqrt{2i a k_z}} \right) \\
		&-\sqrt{\frac{\pi e^{i \omega^2/a k_z}}{2 i a k_z}}
			\text{Erf}\left( \frac{i \omega'}{\sqrt{2i a k_z}} \right);
\end{split}
\end{equation}
which is in general non-zero in contrast to the case of inertial motion.
Therefore, the total excitation probability, defined by,
\begin{equation*}\label{eq:acceleration_effect_UDW}
	P_\t{U} = \frac{q^2 c^2}{8\pi^2} \int_{-1}^{1} dz \int_0^{\infty} dk \, k\,
	\abs{\mathcal{I}}^2 ,
\end{equation*}
where, $z=\cos(\theta)\in[-1,1]$, can be non-zero. 
Similarly, the excitation probability density for the detector to be excited while emitting a
photon of momentum $k$ irrespective of direction, becomes,
\begin{equation*} \label{eq:acceleration_effect_UDW(k)}
	P_\t{U}(k) = \frac{q^2 c^2 k}{8\pi^2} \int_{-1}^{1} dz \, \abs{\mathcal{I}}^2.
\end{equation*}
The symmetry of the problem along the $z-$axis means that the photon emission is azimuthally symmetric, so that
it is useful to consider the probability density, 
\begin{equation*} \label{eq:acceleration_effect_UDW(k,z)}
	P_\t{U}(k,z) = \frac{q^2 c^2 k}{8\pi^2} \, \abs{\mathcal{I}}^2
\end{equation*}
corresponding to a photon of momentum $k$ emitted along the polar angle, $\theta = \cos^{-1} z$.

\begin{figure}[t!]
	\centering
	\includegraphics[width=0.8\columnwidth]{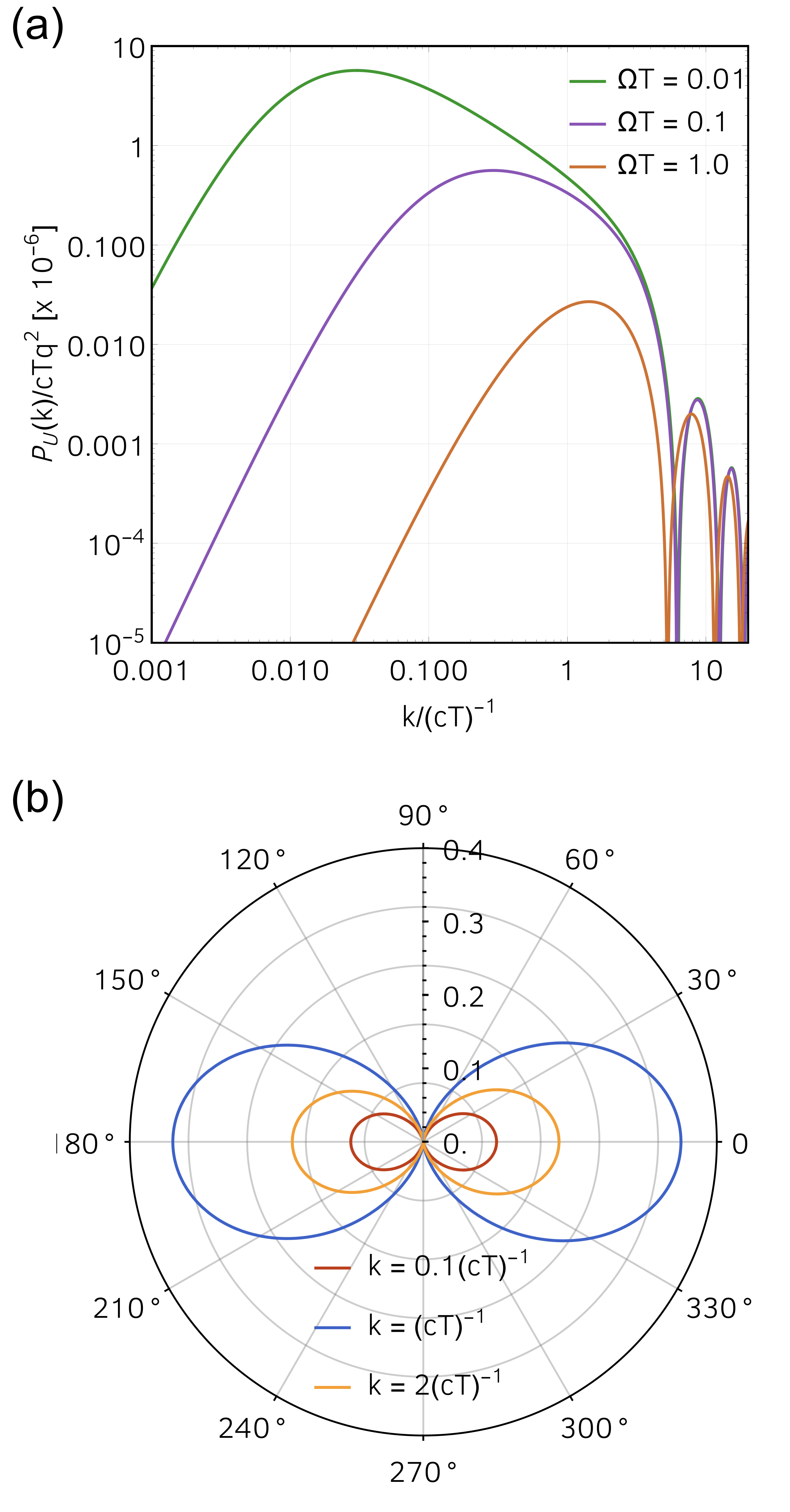}
	\caption{\label{fig1}
	Emission probability for an accelerated UDW detector. (a) The total (i.e. angle-integrated) probability
	$P_\t{U}(k)$ for the emission of an Unruh quantum with momentum $k$, for various values of the
	energy gap $\Omega$.
	(b) Angle-resolved emission probability $P_\t{U}(k,\theta)$ (in units of $10^{-6} cT q^2$), for energy gap
	$\Omega = 0.2/T$.
	Both plots are for acceleration, $a=8\cdot 10^{-3}(c/T)$.
	Note that the oscillations are in \Cref{fig1}a arise from accelerating the detector for a time interval that is compact. 
	}
\end{figure}

\Cref{fig1}a shows the probability $P_\t{U}(k)$ for the excitation of the UDW detector by the emission of
an Unruh quantum of momentum $k$; by momentum conservation, this is equivalent to the angle-integrated probability
of finding an Unruh quantum of momentum $k$.
Note that for the case we consider here, where the detector is not eternally accelerated, the emission is not
isotropic (see \cref{fig1}b), precluding complete analogy with blackbody radiation. 
In fact, radiation is preferentially emitted along/against the direction of acceleration. 
(For realistic scenarios involving charges or atoms as UDW detectors, the 
correct vacuum would not be that of the scalar quantum field --- which is what we consider here --- but the
vacuum of the full vector electromagnetic field; the vectorial character of the latter is expected to produce
radiation transverse to the acceleration, in analogy with classical synchrotron 
radiation \cite{jackson_classical_1999}.)
However there is one aspect of blackbody radiation that is reflected in \cref{fig1}a:
as the energy gap gets smaller the peak of the emission shifts to lower momenta, 
an observation that can be put in correspondence with Wien's displacement law for blackbody radiation 
(i.e. $k_\t{peak}\propto \t{temperature}$) if we associate a temperature proportional to the energy gap
of the detector. (Note that one can always assign a temperature for a two-level system whenever its density 
matrix is diagonal: the ratio of the diagonal elements can be compared to that of the canonical thermal state, and so an 
effective temperature can be defined.)
In this sense, we may formally associate a temperature to the Unruh process, 
even in the case where the detector is accelerated only for a finite time interval. 
Finally, in the non-relativistic regime, when the emitted photon momentum
is ``small'', we have that, $\beta -\gamma = aT k_z \ll 1$ (clearly, ``small'' means, $k_z \ll 1$, which, in
the dimensioned units of \cref{fig1}a reads, $k_z \ll (c T)^{-1}$); explicitly expanding
the amplitude integral in \cref{eq:I_Unruh_1} in the small parameter $\beta-\gamma \propto a$, one can
also show that $P_\t{U}\propto a$. Thus we are able to formally establish that the temperature is proportional to 
the acceleration even in the case where the detector is accelerated only for a finite duration.

\section{Unruh effect with finite-mass detector}

Experimentally producing the large accelerations required to observe the Unruh effect call for 
low mass UDW detectors.
Any such detector will experience significant recoil once an Unruh quantum is emitted. Once the possibility
of recoil is admitted, it becomes unphysical to consider an externally prescribed acceleration, even for a finite
time interval (especially if we are interested in measuring the random recoil over an ensemble of multiple emission events). 
We also envision the possibility of inferring the Unruh effect through a direct experimental measurement
of the recoil; the required measurement sensitivities are expected to be comparable to the quantum fluctuations
of the detector center-of-mass degree of freedom.
For these reasons, we must treat the detector's center-of-mass in a full quantum framework, and,
self-consistently incorporate external agency that accelerates the system.

To this end, we consider a massive detector with a quantized center-of-mass degree of freedom which couples to 
a quantum scalar field $\hat{\phi}$ \cite{functional_calculus,Stritzelberger_Kempf}. 
The detector's center of mass is coupled to a classical ``electric'' field $\mathbf{E}$, which allows us to 
dynamically model the acceleration of the detector. 
This scenario is modeled by the Hamiltonian,
\begin{align*}\label{eq:electric_field_UDW_Hamiltonian}
	\hat{H} = \frac{\hat{\mathbf{p}}^2}{2M} &- q \, \mathbf{E} \cdot \hat{\mathbf{x}}
    + \Omega \ket{e}\bra{e} 
    + \int d^3 k \, ck \, \hat{a}_{\mathbf{k}}^{\dagger} \hat{a}_{\mathbf{k}}^{\vphantom \dagger} \nonumber \\
    & +q \int d^3 x  \, \hat{\mathcal{P}}(\mathbf{x}) \otimes \hat{\mu} \otimes \hat{\phi}(\mathbf{x})
\end{align*}
with $M$ the mass of the detector, and, $\hat{\mathcal{P}}(\mathbf{x})=\ket{\mathbf{x}}\bra{\mathbf{x}}$ is
the projector onto the center-of-mass position eigenstates.
In the interaction picture, the Hamiltonian reads,
\begin{equation}\label{eq:interaction_Hamiltonian_massive_Unruh}
	H_\t{int}(t) = 
	q
    \int d^3x \,
    \hat{\mathcal{P}}(\mathbf{x},t) \otimes 
    \hat{\mu}(t) 
    \otimes 
    \hat{\phi}(\mathbf{x}, t)
    \,,
\end{equation}
with $ \hat{\mathcal{P}}(\mathbf{x},t) = \ket{\mathbf{x}(t)} \bra{\mathbf{x}(t)}$.
In order to model a scenario comparable to the situation considered in the previous section, 
we assume an electric field,
\begin{equation*}\label{eq:constant_electric_field}
	\mathbf{E}(t) = \mathcal{E} \, \Theta(t) \, \Theta(T-t) \mathbf{e}_z,
\end{equation*}
with a non-zero strength $\mathcal{E}$ in the time interval $t\in[0,T]$,
and zero elsewhere. It models the detector's center-of-mass being 
uniformly accelerated in that interval, while it evolves freely for $t\notin [0,T]$. 

Before we can study the vacuum excitation process for the massive detector, we need the time evolved operators 
$ \hat{\mathcal{P}}(\mathbf{x},t) = \ket{\mathbf{x}(t)} \bra{\mathbf{x}(t)}$. To this end, we write the 
Heisenberg equation for the detector's position and momentum,
\begin{equation*}
	\dot{\hat{\mathbf{x}}}(t) = \frac{\hat{\mathbf{p}}(t)}{M}, \qquad
	\dot{\hat{\mathbf{p}}}(t) = q \mathcal{E}(t) \mathbf{e}_z,
\end{equation*}
which produces the time-dependent position operator,
\begin{equation*}
	\hat{\mathbf{x}}(t) = \hat{\mathbf{x}}(0) + \hat{\mathbf{p}}(0)t/M
    + f(t) \, \mathbf{e}_z,
\end{equation*}
where,
\begin{equation*}
	f(t) = \frac{a}{2}\left[ t^2 \Theta(t)\,\Theta(T-t)
    + T(2t-T) \Theta(t-T) \right],
\end{equation*}
with, $a=q\mathcal{E}/M$, being the uniform acceleration due to the electric field. 
We note that $f(t)$ is of the same form as the $z-$component of the classical trajectory which we prescribed in Eq.(\ref{eq:accelerated_prescribed_trajectory}) for the UDW detector with classical center of mass.
Since the position and momentum operators coincide at time $t=0$ between the Heisenberg and Schrödinger pictures,
we have that, $\hat{\mathbf{x}}(0)\psi(\mathbf{x}) = \mathbf{x} \psi(\mathbf{x})$ and 
$\hat{\mathbf{p}}\,(0)\psi(\mathbf{x}) = -i \mathbf{\nabla} \psi(\mathbf{x})$.
Next, to find the time dependent position eigenfunction $\psi_{\boldsymbol{\xi}\,}(\mathbf{x},t) = \langle \mathbf{x} | \psi_{\boldsymbol{\xi}\,}(t) \rangle$ for a given position eigenvalue $\boldsymbol\xi$, we solve the 
Schrödinger equation,
\begin{equation*}
	\left( \mathbf{x} 
    + f(t) \mathbf{e}_z
    - \frac{it}{M} \nabla \right) \psi_{\boldsymbol{\xi}\,}(\mathbf{x},t)
    = \boldsymbol\xi \, \psi_{\boldsymbol{\xi}\,}(\mathbf{x},t),
\end{equation*}
with the initial condition 
$\ket{\psi_{\boldsymbol{\xi}\,}(0)} = \ket{\boldsymbol{\xi}}$, and enforcing the normalization condition, 
\begin{equation*}
	\int d^3x \, 
    \psi^*_{\boldsymbol{\xi}\,}(\mathbf{x},t)
    \psi_{\boldsymbol{\xi}'\,}(\mathbf{x},t) 
    = \delta^{(3)}(\boldsymbol{\xi}-\boldsymbol{\xi}').
\end{equation*}
The required wavefunction is,
\begin{equation*}
	\ket{\psi_{\boldsymbol{\xi}}(t) } =
	\int\frac{d^3p}{(2\pi)^{3/2}} \, 
	\exp \left[ it \frac{p^2}{2M} -i \mathbf{p}\cdot \mathbf{\xi} +i p_z f(t) \right]
    \ket{\mathbf{p}}.
\end{equation*}
Putting all this together, we obtain the time evolved projection operator,
\begin{align*}
	\hat{\mathcal{P}}(\mathbf{x},t) 
	= \int \frac{d^3p d^3q}{(2\pi)^3}
	& \exp\Big[ it \frac{q^2-p^2}{2M} - i (\mathbf{q}-\mathbf{p})\cdot \mathbf{x}  \\
	&\qquad +i (q_z -p_z)f(t) \Big]
    \ket{\mathbf{q}} \bra{\mathbf{p}},
\end{align*}
that fully determines the interaction hamiltonian in \cref{eq:interaction_Hamiltonian_massive_Unruh}.

\subsection{Transition amplitude, probability and probability densities}\label{sec:massive_transition}

We are now equipped to study what we refer to as the \emph{massive Unruh effect}, that is, the excitation process 
both of a UDW detector with a finite mass, initially in the ground state of its internal degree of freedom coupled 
to a scalar quantum field initially in its vacuum state, and accelerated by an external electric field. 
That is, we consider initial and final states of the form,
\begin{align*}
	\ket{\psi_\t{i}} &= \ket{\varphi} \otimes \ket{g} \otimes \ket{0} \\
	\ket{\psi_\t{f}} &= \ket{\mathbf{r}} \otimes \ket{e} \otimes \hat{a}^{\dagger}_{\mathbf{k}} \ket{0}
\end{align*}
where, $\ket{\varphi}=\int d^3p \, \tilde{\varphi}(\mathbf{p}) \ket{\mathbf{p}}$ is
the initial center of mass state, $\ket{\mathbf{p}}$ the center of mass momentum eigenstates,
$ \tilde{\varphi}(\mathbf{p})$ the initial center of mass wave function in the momentum representation,
and $\mathbf{r}$ the detector's recoil momentum.

The transition amplitude for the process where the detector gets excited, its center-of-mass recoils with momentum 
$\mathbf{r}$, and emits an Unruh quantum of momentum $\mathbf{k}$, is (upto a phase factor),
\begin{equation*}\label{eq:A_massive0}
	\mathcal{A}_\t{M} = 
	\frac{q c}{2\pi\sqrt{4\pi k}} \,
    \tilde{\varphi}(\mathbf{r}+\mathbf{k}) \,
    \mathcal{J}(\mathbf{r}),
\end{equation*}
where we define,
\begin{equation*}\label{eq:JM}
	\mathcal{J}(\mathbf{r}) = \int_{-\infty}^{\infty} dt\,
	\exp \left[ it \left( \frac{k^2}{2M} -\frac{\mathbf{r}\cdot \mathbf{k}}{M} 
		+ ck + \Omega \right) -i k_z f(t) \right].
\end{equation*}
The corresponding transition probability density is,
\begin{equation*}\label{eq:P(k,r)_electric_field_switched}
	P_\t{M}(\mathbf{k},\mathbf{r}) 
	= \abs{\mathcal{A}_\t{M}}^2 = 
	\frac{q^2 c^2}{(2\pi)^2 4\pi k} \,
    \abs{\tilde{\varphi}(\mathbf{r}+\mathbf{k}) \mathcal{J}(\mathbf{r})}^2
\end{equation*}

To study the recoil of the detector, we consider the
excitation probability density as a function of the recoil momentum $\mathbf{r}$ (and irrespective of
the momentum of the emitted photon),
\begin{equation*}
	P_\t{M}(\mathbf{r}) =
		\int d^3 k \, \frac{q^2 c^2}{(2\pi)^2 4\pi k}
    \abs{ \tilde{\varphi}(\mathbf{r}+\mathbf{k}) \mathcal{J}(\mathbf{r}) }^2;
\end{equation*}
and the total excitation probability for the massive Unruh process,
\begin{equation*}\label{eq:P_electric_field_switched}
	P_\t{M} = \int d^3 k \int d^3 p \, \frac{q^2 c^2}{(2\pi)^2 4\pi k} \,
	\abs{\tilde{\varphi}(\mathbf{p}) \mathcal{J}(\mathbf{p}-\mathbf{k})}^2 .
\end{equation*}
To resolve the angular dependence of the emission and recoil, we write $\mathbf{k}=(k_x,k_y,k_z)
= (k\sin(\theta)\cos(\phi) , k\sin(\theta)\sin(\phi) , k\cos(\theta))$, with $\phi$ the azimuthal angle 
and with $\theta$ the polar angle, that is, the angle between the momentum of the emitted photon and the direction of 
the electric field lines (and therefore, of the acceleration);
as before, we also define, $z=\cos(\theta)$ 
(in the following we will refer to both $z$ and $\theta$ as the polar angle of the emitted photon).
The excitation probability density for the process to happen while emitting an Unruh quantum of momentum $k$ in
magnitude, irrespective of direction, is
\begin{equation}\label{eq:P(k)_electric_field_switched}
\begin{split}
	P_\t{M}(k) &= 
		\int_{-1}^{1} dz
 		\int_0^{2\pi} d\phi 
    	\int d^3 k \int d^3 p \,
    	\frac{q^2 c^2 k}{(2\pi)^2 4\pi} \\
    &\qquad\times 
    \abs{\tilde{\varphi}(\mathbf{p}) \mathcal{J}(\mathbf{p}-\mathbf{k})}^2
\end{split}
\end{equation}
Similarly, we define the excitation probability density as a function of both the magnitude $k$ and polar angle 
$z$ of the emitted photon:
\begin{equation}\label{eq:P(k,z)_electric_field_switched}
	P_\t{M}(k,z) = \int_0^{2\pi} d\phi 
    \int d^3 p \,
    \frac{q^2 c^2 k}{(2\pi)^2 4\pi} \abs{\tilde{\varphi}(\mathbf{p})\mathcal{J}(\mathbf{p}-\mathbf{k})}^2
\end{equation}
All the above expressions depend on the time integral $\mathcal{J}$, which can be explicitly evaluated,
\begin{equation}\label{eq:Jpk}
\begin{split}
	\mathcal{J}(\mathbf{p}-\mathbf{k}) 
    = \frac{1}{i \omega_\t{M}} &- \frac{e^{iT(\omega_\t{M}'+ak_z T/2)}}{i \omega_\t{M}'} \\
    	&+ \sqrt{\frac{\pi e^{i \omega_\t{M}^2/a k_z}}{2i a k_z}}  
    		\text{Erf}\left( \frac{i \omega_\t{M}}{\sqrt{2i a k_z}} \right) \\
    	&- \sqrt{\frac{\pi e^{i \omega_\t{M}^2/a k_z}}{2i a k_z}}
    		\text{Erf}\left( \frac{i \omega_\t{M}'}{\sqrt{2i a k_z}} \right)
\end{split}
\end{equation}
in terms of, $\omega_\t{M} = \omega +\frac{k^2}{2M} - \frac{\mathbf{p} \cdot\mathbf{k}}{M}$, and,
$\omega_\t{M}' = \omega_\t{M} - ak_z T$.
In writing the above expression,
we have omitted two terms each involving delta distributions $\delta(\omega_\t{M})$ and $\delta(\omega_\t{M}')$, 
which is justified for the following reasons. 
First, let us write 
$\mathbf{p}\cdot \mathbf{k}=pk \cos(\kappa)$, 
with $\kappa$ the angle between $\mathbf{p}$ and $\mathbf{k}$, and define,
$\omega_0 = \omega + \frac{k^2}{2M}$ for brevity. The delta distribution $\delta(\omega_\t{M})$ 
then peaks only for 
$\cos(\kappa)=\frac{M \omega_0}{pk}$. 
Furthermore, the delta distribution 
$\delta(\omega_\t{M}')$ 
peaks only for 
$\cos(\kappa)=\frac{M}{pk}(\omega_0 - aT k_z )$. 
But since 
$\cos(\kappa)\in[-1,1]$, 
a necessary condition for $\delta(\omega_\t{M})$ to peak is $p\geq Mc $, which translates to saying that the initial virtual center of mass velocities would have to be superluminal, which is ruled out.  
Similarly, a necessary condition for $\delta(\omega_\t{M}')$ to peak is $p+MaT\geq Mc$, which would require the virtual center of mass velocities to be superluminal by the end of the accelerated phase. 
Physically, the delta distributions $\delta(\omega_\t{M})$ and $\delta(\omega_\t{M}')$ have their origin in the virtual inertial motion of the detector, respectively for the times $t< 0$ and $t>T$ during which the electric field is switched off. Inertial virtual motion (just like inertial real motion) should not cause excitation of the detector and the field, which is reflected in the vanishing of these delta distributions.

\subsection{The Unruh effect as a limiting case of the massive Unruh effect}

Before proceeding further, let us see how to recover the conventional Unruh effect (of \cref{Sec:Unruh_effect_UDW}) --- 
i.e. a UDW detector with a prescribed classical trajectory --- 
from the ``massive Unruh effect" studied above.

In order to recover the traditional Unruh effect for a detector experiencing a uniform acceleration $a$, let us 
consider the limit of infinite detector mass --- so that the center of mass wave function delocalizes infinitely 
slowly and so it effectively behave classically. 
A classical particle of charge $q$ and mass $M$ in a constant electric field $\mathcal{E}$ experiences an acceleration 
$a=q\mathcal{E}/M$. Defining $M=m\gamma$ and $\mathcal{E}=\varepsilon\gamma$ and letting $\gamma\rightarrow\infty$ allows us to keep the acceleration $a$ experienced by the detector constant, while considering the infinite mass limit:
\begin{equation*}
	\lim_{\gamma\rightarrow \infty} P_\t{M} = P_\t{U}.
\end{equation*}
The above equation holds true for all the probabilities defined above; in this sense, the massive Unruh effect 
subsumes the conventional (infinite-mass) Unruh effect.

\subsection{Example of a Gaussian center of mass wave packet}

In order to apply the formalism developed in \cref{sec:massive_transition} for the massive Unruh effect,
we need to specify an initial wave function, $\tilde{\varphi}(\mathbf{p})$, for the detector's center of mass.
We consider a Gaussian initial center of mass (momentum) wave packet of the form,
\begin{equation*}\label{eq:initial_momentum_wavepacket}
	\tilde{\varphi}(\mathbf{p}) = \left( \frac{L^2}{2\pi}\right)^{3/4} e^{-p^2L^2/4},
\end{equation*}
assumed to exist at time $t=0$, so that when the electric field is switched on, the detector's center of mass
in position space is localized at the origin with width $L$.

Since we work in the non-relativistic regime, it is necessary to choose the parameters $L$, $M$, $T$ and $a$ 
in a way that ensures that the detector's virtual center of mass velocities are much less than the speed of light. 
The initial momentum of the detector in the $z-$direction, i.e., parallel to the electric field, 
is Gaussian distributed around $p_z=0$, with a standard deviation of $\sqrt{2}/L$. 
Initial momenta which are $\sigma$ standard deviations away from the mean then correspond to initial virtual 
center of mass speed, $\abs{v_z(0)} = 2\sigma \cdot \sqrt{2}/(LM)$. 
If we require that these tails of the wavepacket, after having the accelerating field $\mathcal{E}$ turned on 
for a time $T$, be still less than $1\%$ the speed of light, then we need to maintain an electric field
that satisfies, $v_z(T) = \abs{v_z(0)} + a T \lesssim 0.01 c$. In the following, we choose electric field
strengths weak enough that this is satisfied for virtual velocities that are $\sigma = 3.5$ standard deviations
from the mean velocity.
This means that the center-of-mass velocity is a small fraction of the speed of light, i.e. $v/c = p/(Mc) \ll 1$; 
thus $\abs{\mathcal{J}(\mathbf{p}-\mathbf{k})}^2$ (from \cref{eq:Jpk}) can be approximated around 
$\mathbf{p}/(Mc)=0$ by using only its first two Taylor coefficients:
\begin{align*}
	J_1 &= \left[ \abs{\mathcal{J}(\mathbf{p}-\mathbf{k})}^2 \right]_{\omega = \omega_0}, \\
	J_2 &= \frac{1}{2}\left[ \frac{\partial^2}{\partial A^2} \abs{\mathcal{J}(\mathbf{p}-\mathbf{k})}^2 
		\right]_{\omega = \omega_0}.
\end{align*}
Expressing $J_{1,2}$ in terms of the variables $k$ and $z$, we then obtain for the 
basic excitation probability densities introduced above (valid to $\mathcal{O}( (LMc)^{-4} )$):
\begin{equation*}\label{eq:P_Massive_Gaussian}
\begin{split}
	P_\t{M}(k,z) &\approx \frac{q^2 c^2 k}{8\pi^2}\left[ J_1(k,z) + \frac{k^2 J_2(k,z)}{(ML)^2} \right] \\
	P_\t{M}(r,\zeta) &\approx \frac{L^3 q^2 c^2}{2(2\pi)^{5/2}} \int_0^\infty dk \int_{-1}^{1}dz\, 
		k e^{-(r^2 + k^2 +2rk \zeta) L^2/2} \\
		&\quad \times \left[ J_1(k,z) 
			+ \frac{(rk\zeta +k^2)^2}{M^2}  J_2(k,z) \right]
\end{split}
\end{equation*}
Here, $r$ is the magnitude of the recoil momentum and $\zeta = \mathbf{r}\cdot \mathbf{k}/(rk)$
is the cosine of the angle between the recoil momentum ($\mathbf{r}$) and Unruh photon momentum ($\mathbf{k}$).

\begin{figure}[t!]
	\centering
	\includegraphics[width=0.8\columnwidth]{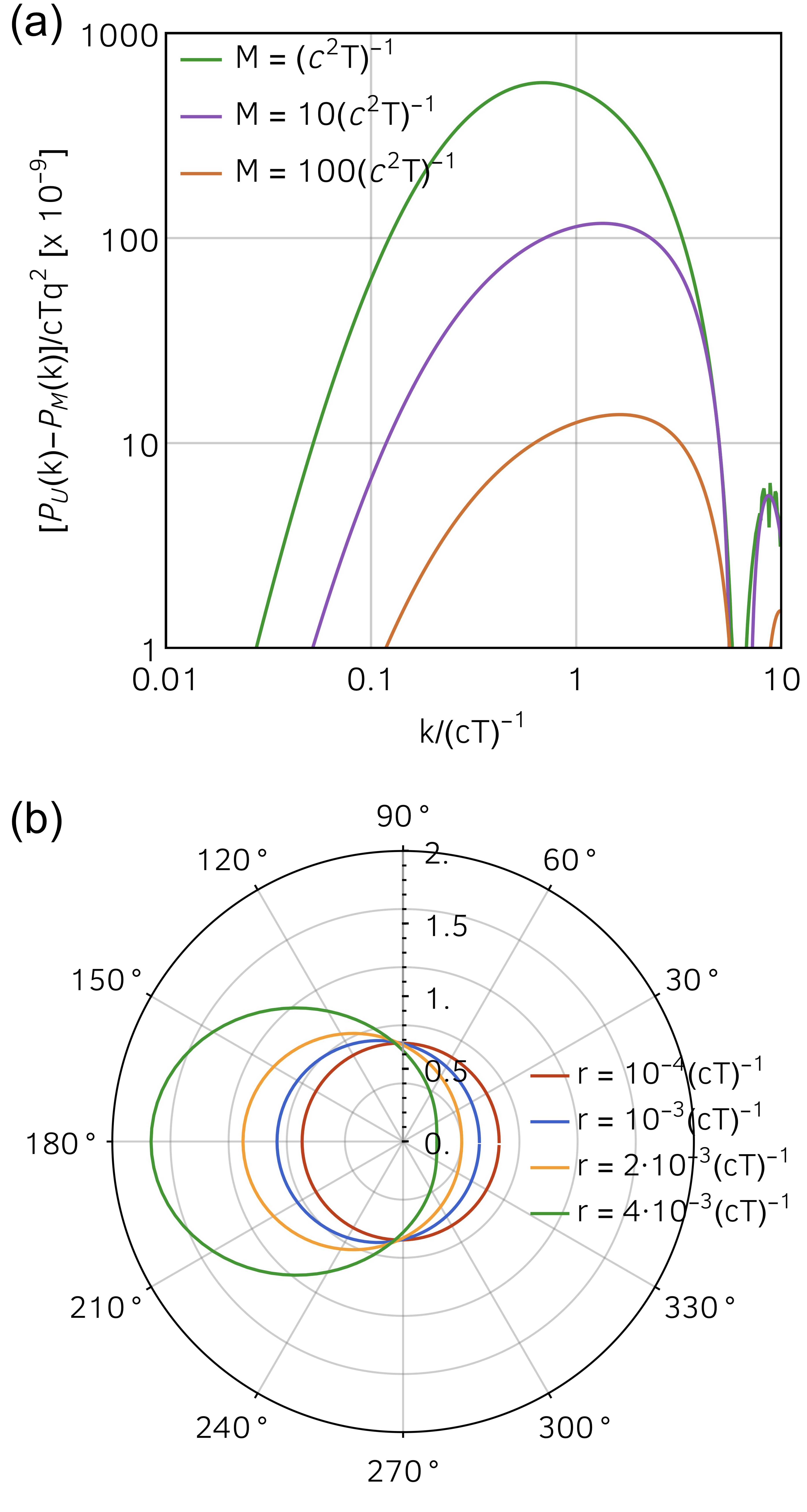}
	\caption{\label{fig2}
	Emission and recoil probability for a massive UDW detector. 
	(a) Plot shows the difference of the emission
	probabilities between the cases where the detector center-of-mass is infinitely heavy (i.e. $P_\t{U}(k)$)
	and the case where the detector center-of-mass has a finite mass (i.e. $P_\t{M}(k)$).
	(b) Recoil probability density $P_\t{M}(r,\zeta)$ (in units of $10^{-6}(cTq^2)^{-1}$) as a function of
    recoil momentum angle $\zeta$, for a few values of the magnitude of the recoil momentum magnitude $r$.
    The detector mass is $M=10/(c^2 T)$, while in both panels, it's internal energy gap is $\Omega = 0.2/T$, 
    acceleration $a = 8\cdot 10^{-3}(c/T)$, and the center-of-mass wavepacket has an initial width, $L=100(cT)$. 
	}
\end{figure}

\Cref{fig2}a depicts the difference, $P_\t{U}(k)-P_\t{M}(k)$, between the angle-integrated emission probabilities
of the conventional infinite-mass UDW detector and a UDW detector of finite mass, as a function of
the momentum $k$ of the Unruh quantum. As the mass increases, 
the finite mass case approaches the standard case (i.e. $P_\t{U}-P_\t{M}\rightarrow 0$) as expected. 

Although the total emission probability, and hence the emitted flux of Unruh quanta, is seen to decrease with mass, 
each individual emission causes greater recoil of the detector center-of-mass for a lower mass detector. 
This is shown in \cref{fig2}b, where the angular recoil probability is plotted
as a function of the magnitude of the recoil momentum. Recoil with larger momenta happen in the direction opposite
to the acceleration ($0^\circ$ in the polar plot) --- a consequence of momentum conservation.

\section{Conclusion}

We analyzed the behavior of UDW detectors which possess a quantum mechanical center of mass and that are  finitely accelerated through the vacuum of a scalar quantum field. We found a characteristic interplay between the acceleration-induced excitation of the UDW detector, the non-isotropic patterns of the flux of the emitted Unruh quanta and the corresponding quantum recoil. This makes the quantum recoil a potentially experimentally relevant signature of the Unruh effect. 

In practice, trapped and accelerated (sub-)atomic particles with internal states can act as UDW detectors. The ability to precisely measure forces that act on them may then provide a route to a direct detection of the
Unruh effect on a single particle. 
Indeed, electron bunches in storage rings have long been suspected of being probes of the 
Unruh effect \cite{BellLein87,Schutz06}. However, space charge effects and other systematics have prevented a 
decisive measurement of the Unruh effect (nor does the accelerator environment offer an ideal venue to
study more fundamental predictions of the Unruh effect).

Although any realistic experiment along these lines will have to consider the coupling of the internal degree of freedom
to the electromagnetic vacuum, the scalar field scenario considered here allows a first qualitative glimpse of what to expect.
Further studies along these directions are in progress. 

\bibliographystyle{apsrev4-1}
\bibliography{references}

\end{document}